\documentclass[usenatbib,fleqn]{mn2e}

\newif\ifpdf
\ifx\pdfoutput\undefined
   \pdffalse              
\else
   \pdfoutput=1           
   \pdftrue
\fi
\ifpdf
  \usepackage[pdftex]{graphicx}
  \usepackage[bookmarks=true]{hyperref}
  \hypersetup{%
        bookmarksopen=true,
        colorlinks=true,
        pdfpagemode=UseOutlines,
        pdftitle={Orientation and size of the `Z' in XRGs},
        pdfsubject={AGN, jets, merger of galaxies},
        pdfauthor={Dr.~C.~Zier}
  }
  \usepackage{thumbpdf}
\else
  \usepackage[dvips]{graphicx}
\fi

\usepackage{xspace}
\usepackage{latexsym}
\usepackage{amssymb}
\usepackage{amsmath}
\usepackage{journals}

\newcommand{\eqn}[1]{Eq.~(\ref{#1})}
\newcommand{\fgr}[1]{Fig.~\ref{#1}}
\newcommand{\mrm}{\mathrm}
\newcommand{\mbf}{\mathbf}
\newcommand{\bvec}[1]{\mbox{\boldmath${\mathrm{#1}}$\unboldmath}}
\newcommand{\HI}{H\,{\sc i}\xspace}
\newcommand{\X}{\textsf{X}\xspace}
\newcommand{\Z}{\textsf{Z}\xspace}
\newcommand{\BH}{SMBH\xspace}

\def\beq{\begin{equation}}
\def\eeq{\end{equation}}
\def\beqa{\begin{eqnarray}}
\def\eeqa{\end{eqnarray}}

\defcitealias{gopal-krishna03}{\mbox{G-KBW}}

\title[Orientation and size of the `\Z' in XRGs]{Orientation and size of the
  `\Z' in \X-shaped radio galaxies}

\author[C. Zier]{C. Zier$^{1}$\thanks{E-mail: chzier@rri.res.in}\\
$^{1}$Raman Research Institute, Bangalore 560080, India}
\begin{document}

\date{Accepted Received}

\pagerange{\pageref{firstpage}--\pageref{lastpage}} \pubyear{2005}

\maketitle

\label{firstpage}

\begin{abstract}
Some \X-shaped radio galaxies show a \Z-symmetric morphology in the less
luminous secondary lobes. Within the scenario of a merger between two
galaxies, each hosting a supermassive black hole in its center, this structure
has been explained before: As the smaller galaxy spirals towards the common
center, it releases gas to the ISM of the larger active galaxy. The ram
pressure of this streaming gas will bend the lobes of the pre-merger jet into
a \Z-shape. After the black holes have merged the jet propagates in a new
direction that is aligned with the angular momentum of the binary black
hole. In this article we deproject the pre- and post-merger jets. Taking into
account the expected angles between the jet pairs and with the assumption that
their directions are uncorrelated, we show that one of three possible
orientations of the jets with respect to the line of sight is more likely than
the others. This actually depends on the distance where the bending occurs.
Another result of our deprojection is that the streaming gas bends the jet
into a \Z-shape in a range between about $30$ and $100\,\mrm{kpc}$ distance to
the center of the primary galaxy. We confirm this finding by comparing our
predictions for the properties of the rotational velocity field and its radius
with observations and numerical simulations of merging galaxies. Thus our
results support the merger scenario as explanation for \X- and \Z-shaped radio
galaxies with the jet pointing along the former axis of orbital angular
momentum of the binary.
\end{abstract}

\begin{keywords}
galaxies: active -- galaxies: interactions -- galaxies: jets -- galaxies:
kinematics and dynamics -- galaxies: individual: NGC~326 -- galaxies:
individual: 3C~52
\end{keywords}

\section{Introduction}
\label{s_intro}
A small subclass of radio galaxies is formed by the \X-shaped radio galaxies
(XRGs), which all have radio luminosities close to the FR~I/II transition of
$10^{25}\,\mrm{W/Hz}$ at $178\,\mrm{MHz}$ \citep{fanaroff74}. These sources
show two misaligned pairs of radio lobes of comparable extent
\citep[e.g.][]{ekers78,leahy92}, which have also been referred to as wings and
appear as \X-shaped structures. After the discovery of this class of sources
various mechanisms for their formation have been proposed. Here we give only a
short summary, for a more detailed discussion see e.g. \citet{rottmann01} or
\citet{dennett-thorpe02}.

According to the backflow model by \citet{leahy84} jet material is streaming
from the hot spots of the primary lobes back towards the host galaxy. It
remains collimated until it hits the backflow from the opposite lobe and then
expands laterally in a fat disk perpendicular to the radio lobes. It is not
clear in this model how the plasma, falling back on the disk, can be diverted
to just one side of the primary lobes in order to form the \X-shape. Moreover
this mechanism can not explain that the secondary lobes extend as far or even
farther than the primary lobes and how such an extension can be achieved with
subsonic velocities within the life-time of the radio source.

The buoyancy model suggests that the radio lobes have a lower density than the
ambient medium, resulting in buoyant forces \citep{gull73,cowie75}, which are
thought to bend the lobes towards regions that provide density
equilibrium. This model has the same problems as the backflow model to explain
the symmetry of XRGs and the extension of the secondary lobes.

\begin{figure*}
\ifpdf
  \includegraphics[width=84mm]{ngc_326.png}\hfill
  \includegraphics[width=72mm]{3C_52.png}
\else
  \includegraphics[width=84mm]{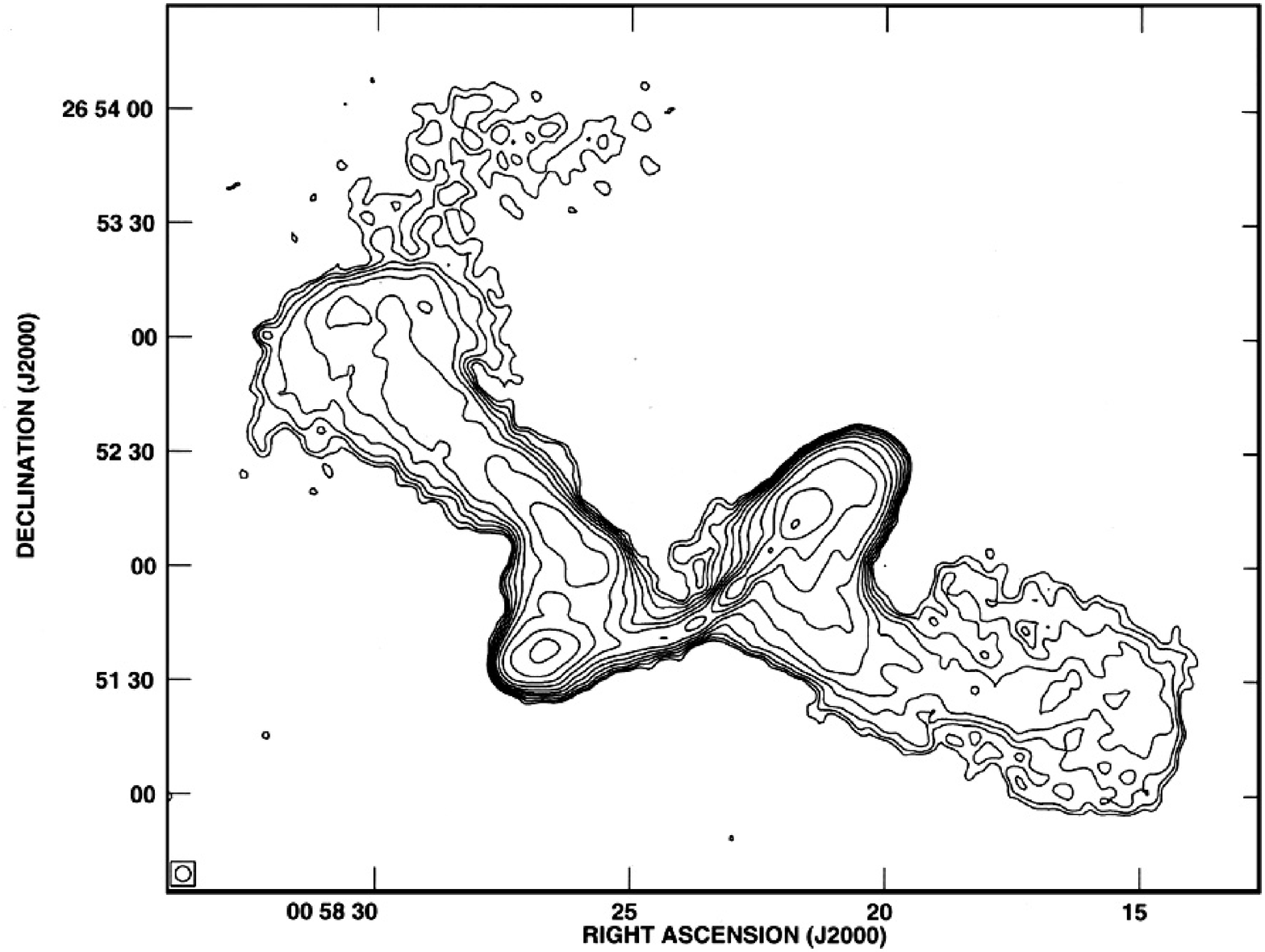}\hfill
  \includegraphics[width=72mm]{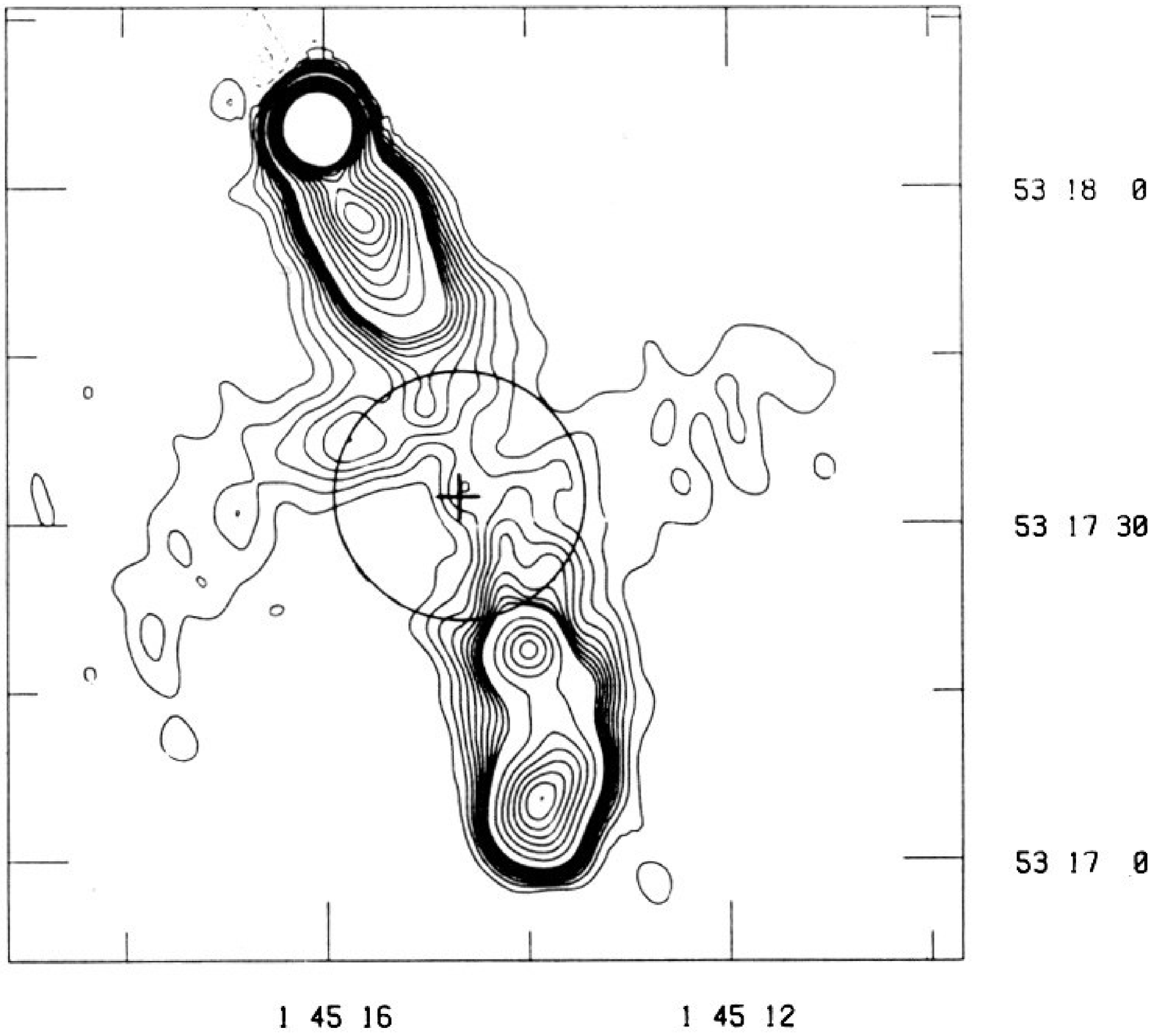}
\fi
\caption[]{Radio images of the two \Z-shaped XRGs are shown. NGC~326
  \citep{murgia01} in the left panel at $1.4\,\mrm{GHz}$ and 3C~52
  \citep{leahy84} in the right at $1.6\,\mrm{GHz}$. In both sources the
  secondary lobes are weaker and their ridges have a lateral offset of about
  their width, forming a \Z-shape. While in 3C~52 primary and secondary lobes
  have about the same extension, the secondary ones in NGC~326 are much more
  extended.}
\label{f_zrgs} 
\end{figure*}

Apparently the radio jets, which are assumed to be aligned with the spin of
the black hole at their origin \citep{wilson95}, have been stable for a long
time span before undergoing a short period of reorientation into another
stable state \citep{rottmann01}. This can not be explained by a steady
precession as has been originally done by \citet{ekers78}. \citet{gower82}
applied a precessing jet model to various galaxies, among them NGC~326. Though
the wings could be reproduced the inner symmetry is rotated by about
$45^\circ$. Also according to present day deep radio images of NGC~326 a
precessing jet seems to be highly unlikely the cause for its shape
\citep{rottmann01}. An interesting idea, pointed out by the referee, is
whether such a structure could be related to the combination of a precessing
jet and the recoil suffered by the merged black hole due to anisotropic
emission of gravitational waves (e.g. \citealt{blandford79}). Without going
into the details, what would be beyond the scope of this article, it does not
seem to be very likely for both the sources we are interested in here. Such a
kick would give the black hole, i.e. the source of the post-merger jet, a
linear momentum and hence break the symmetry. Though this might explain the
apparent curve close to the center in the post-merger jet of NGC~326,
generally both jet pairs are seen to be symmetrical about the common
center. The symmetry can rather be explained by rapid realignment of the jet
due to accretion from a misaligned disk or the coalescence of a supermassive
binary black hole (BBH) on time scales less than $10^7\,\mrm{yr}$
\citep{rottmann01,zier01,zier02}. Such a BBH is formed previously in a merger
of two galaxies each of which host one of the supermassive black holes (SMBH)
in their center \citep{begelman80}. The final stage of the merger is dominated
by emission of gravitational radiation which leaves the spin of the resulting
\BH aligned with the orbital angular momentum $\mbf{L}_\mrm{orb}$ of the
binary \citep{rottmann01,zier01,zier02,chirvasa02,biermann02}. After a short
time a new jet will propagate along the spin axis, i.e. the jet flips from the
direction of the spin of the pre-merger \BH in the direction of
$\mbf{L}_\mrm{orb}$ (see Fig.~9 in \citealt{zier02}). These ideas were also
explored by \citet{merritt02}.

In at least two XRGs the ridges of the secondary lobes have been observed to
be offset from each other laterally by about their width, hence showing a
\Z-shaped symmetry about the nucleus (\fgr{f_zrgs}). Because a high angular
resolution as well as a rather special aspect angle are necessary to see such
an offset it is possible that a \Z-symmetry is more common than current
detections suggest. To explain this symmetry \citet{gopal-krishna03}
(hereafter \citetalias{gopal-krishna03}) propose some modifications to the
spin-flip model outlined above: As the captured galaxy spirals to the common
center it induces a rotational stream-field in the ambient medium on large
scales. If the trajectory of the secondary galaxy passes through the polar
regions of the primary, the motion of the ISM bends the original jet into
\Z-shape before the merger is completed. This means that purely \Z-symmetric
radio sources (i.e. without \X-morphology) might be spotted before evolving
to an XRG once the \BH{}s have coalesced.

In the present article we will deproject the \Z-shaped sources in order to
understand their geometry and possible orientation to us. Of interest is the
distance in which the old jet is bent into \Z-shape since it gives us
information about the strength of the jet and the properties of the gas stream
in the wake of the secondary galaxy. Ultimately this contributes to our
knowledge of the history of a merger between two galaxies.

In the next section we will explain the geometry of the jets and lobes and
deduce limits for the involved angles before deriving the expressions for
deprojecting the jets. In Sect. \ref{s_appben} we apply our model to the two
\Z-shaped sources NGC~326 and 3C~52. The results are discussed and compared
with other observations in Sect. \ref{s_discussion}. A summary and conclusions
are presented in Sect. \ref{s_final}.

\section{Jet orientations and \Z-shapes}
\label{s_theory}
\subsection{The geometry of ZRGs}
\label{s_orient}

In order to observe a source as XRG both, the primary and secondary lobes,
i.e. the post- and pre-merger lobes respectively, have to be close to the
plane of sky. Both pairs also have to subtend a sufficiently large angle on
the sky so that we can distinguish them. Because the pre-merger spin of the
primary \BH and the orbital angular momentum of the merging binary, defining
the later post-merger \BH spin, are not correlated, we expect the angle
between them to be large on average. In his thesis \citet{rottmann01} used
statistical methods to estimate the most likely distribution of the intrinsic
angle $\theta$ between both pairs of lobes. For this purpose he constructed a
theoretical distribution of projected angles $\theta'$ for an ensemble of XRGs
with intrinsinc angles uniformly distributed between $\theta_1$ and
$\theta_2$. Taking into account various selection effects (the projected angle
should not be too small or large; inclination of the jets with respect to the
plane of sky should be small; the projected length of the secondary lobe
should not be too short) and comparing the theoretical results with the
observed distribution he obtains the best fit if $60^\circ \lesssim \theta
\lesssim 90^\circ$. This is in agreement with our expectation of the intrinsic
angle to be large. However, \citeauthor{rottmann01} points out that the
obtained distribution can not reproduce the peak observed at $\sim 50^\circ$
in the distribution of the projected angle. He suggests that a non-uniform
distribution of intrinsic angles will improve the fit.

Provided that the direction of the spin of the post-merger \BH and hence also
the direction of the post-merger jet are dominated by the orbital angular
momentum of the binary, the directions of the pre- and post-merger jets are
uncorrelated. This allows us to imagine both jets as a pair of uncorrelated
arrows which we can superpose so that both their centers lie on each other and
they enclose an angle in the range $0^\circ\le\theta\le 180^\circ$. Without
loss of generality we fix one arrow to be aligned with the $z$-axis. Asking
now for the distribution of the intrinsic angles between both arrows is like
looking for the distribution of the pinholes the second arrow pierces through
the surface of the unit-sphere. Since both arrows are uncorrelated this is
analogue to a uniform distribution of stars projected on the
unit-sphere. Therefore the probability to find a star or pinhole in a solid
angle element $d\Omega = \sin\theta\,d\theta\,d\phi$ around the coordinates
$(\theta, \phi)$ is
\[
p(\theta, \phi) = 
\begin{cases}
\frac{d\Omega}{4 \pi} & 0\le\theta\le\pi \quad\text{and}\quad
0\le\phi\le 2\pi,\\
0 & \text{otherwise}.
\end{cases}
\]
With the substitution $u=-\cos\theta$ a uniform distribution over the
unit-sphere requires a uniform distribution of both coordinates in the ranges
$-1\le u\le 1$ and $0\le\phi\le 2\pi$. Hence the intrinsic angle $\theta$,
instead of being uniformly distributed, is distributed according to $p(\theta)
= 1/2\,\sin\theta$, peaking at $90^\circ$. The less the orbital angular
momentum of the binary dominates the post-merger spin the more the maximum in
the distribution of $\theta$ will shift to smaller angles and the more the
distribution will deviate from a symmetric distribution about the maximum.
Qualitatively this seems to be in good agreement with the observed
distribution that \citeauthor{rottmann01} shows in his thesis. A careful
comparision of the theoretical with the observed distribution could give a
clue about the nature of the formation of XRGs and which component dominates
the post-merger spin after the \BH{}s have merged due to emission of
gravitational radiation. But this requires much more and better data. We just
keep the result in mind that the formation mechanism of XRGs is likely to
create jet-pairs with large intrinsic angles, while both lobes have to be
close to the plane of sky so that we can actually observe the \X-shape.

As the secondary galaxy is spiraling into the common center of mass, it will
generate a streaming motion in the merger plane due to mass loss and dragging
along ISM of the primary galaxy. On large distances the density and velocity
of the streaming motion will probably be strong enough to bend the jets into
wings. As the the secondary galaxy spirals inwards to smaller distances the
power of the jet will become stronger and beyond a certain distance no bending
of the jet will be possible. In the following we will refer to the distance
where the bending happens as $r$.  The rotation stream will have some
thickness $2h$ perpendicular to the merger plane and in the distance $r$ be
roughly confined to the surface of a cylinder, which is aligned with the
orbital angular momentum of the galaxies.  The possible orientations of both
jet-pairs and the line of sight (LOS) are shown in \fgr{f_z-shapes}. The
$z$-axis is aligned with the orbital angular momentum and hence identified
with the post-merger jet. With the bold solid circle we denote the merger
plane and with the pair of thin solid and dashed circles we mark the rotation
stream with radius $r$ and minimum half-height $h$ for the three possible
secondary pairs of jets. The LOS is perpendicular to the paper plane and
pointing straight to the center. It encloses an angle $\delta_\mrm{LOS}$ with
the merger plane and $\theta_\mrm{LOS}$ with the $z$-axis so that
$\theta_\mrm{LOS} + \delta_\mrm{LOS} = 90^\circ$. If we would move around the
$z$-axis with constant $\theta_\mrm{LOS}$ our LOS would intersect with the
cylinder of the rotation stream along the dotted circle. The three possible
orientations (\textbf{a}, \textbf{b}, \textbf{c}) of the pre-merger jet
relative to the post-merger jet and the LOS are depicted by the bold lines
with the wings on the circles following the rotational motion after they have
been bent around (here shown in clockwise direction). $\theta_\mrm{jet}$ and
$\delta_\mrm{jet}$ are the intrinsic angle between both jet pairs and the
angle from the pre-merger jet to the merger plane respectively. $\varphi$
denotes the angle between the planes defined by both jet-pairs on the one hand
and the LOS and the $z$-axis on the other hand and therefore lies in a plane
parallel to the merger plane.

\begin{figure}
\ifpdf \includegraphics[width=84mm]{z-shapes.pdf} \else
  \includegraphics[width=84mm]{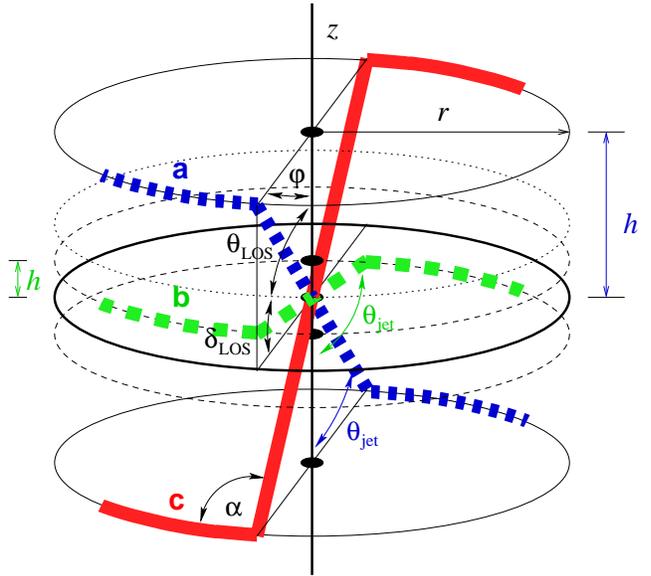} \fi
\caption[]{Possible orientations (\textbf{a, b, c}) between primary lobes
  (aligned with the $z$-axis), the LOS ($\perp$ to the paper plane through the
  center) and the secondary lobes (in color or grey-shades) are
  shown. $\theta$ denotes angles to the $z$-axis, $\delta$ to the merger
  plane, so that $\theta + \delta = 90^\circ$. The secondary lobes are bent in
  the distance $r$ by the clockwise rotating gas into the wings (following the
  solid circle (\textbf{a, c}) or the dashed one (\textbf{b})). $h$ is the
  required minimum half-height of the rotating stream.}
\label{f_z-shapes} 
\end{figure}

\begin{figure*}
\ifpdf
  \includegraphics[width=120mm]{angle-sketch_a.pdf}
\else
  \includegraphics[width=120mm]{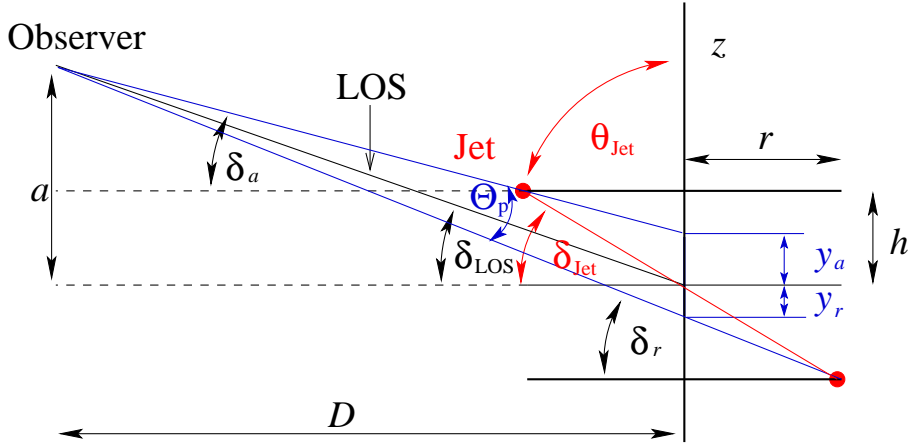}
\fi
\caption[]{Projection of \fgr{f_z-shapes} from the side for orientation
  \textbf{a} with the same meaning of the symbols. $D$ is the observers
  distance to the source, $a$ the vertical distance to the merger plane. The
  angular offset of the ridges of the wings projected in the plane of sky is
  $\Theta_\mrm{p}$. $\delta_a$ and $\delta_r$ denote the angle between the
  merger plane and the line connecting the observer with the bending point of
  the approaching and receding jet respectively. $y_a$ and $y_r$ are the
  offsets of the approaching and receding lobes respectively, projected on the
  $z$-axis.}
\label{f_angle-a} 
\end{figure*}

If we assume the hight of the cylinder, i.e. the streaming matter, to be of
the order of its radius or less, the trajectory of the secondary galaxy
necessarily has to pass through the polar regions of the primary in order to
bend the jet into \Z-shape. Therefore the pre-merger jet and the orbital
angular momentum ($\mbf{L}_\mrm{orb}$), i.e. the post-merger jet, will be
roughly perpendicular to each other. This is consistent with our prediction
for the intrinsic angle above and with the explanation of the \Z-shape within
the framework of the merger model. Hitting the rotation stream the pre-merger
jet will be deflected into the wings at some angle which is determined by the
relative strengths of the ram pressure of the rotating matter and the jet. For
a weak jet the wings will be dragged along with the rotation stream and hence
their projected extension would be limited by the radius of the stream. For
NGC~326 and 3C~52 extensions of at least $50$ and $100\,\mrm{kpc}$ have been
observed respectively. So we can conlude that either the bending happens on
such large radii, or a stronger jet is deflected at a smaller
angle. \fgr{f_zrgs} shows that both wings are almost perpendicular to the
post-merger jet and because their brightness does not differ very much both
wings lie close the plane of sky. Now, a strong pre-merger jet would be
deflected at a small angle and thus it would propagate at an angle a little
larger than $\varphi$ in projection on the merger plane. Because we observe
both wings close to the plane of sky this means that the initial angle
$\varphi$ must have been quite large and as a consequence we should be able to
distinguish the pre-merger jet before the bending from the post-merger jet.
But this has not been observed ($\varphi$ is small) and so we conclude that
large radii of the rotating stream are more likely than small deflecting
angles. Because $\varphi$ is small as well as the angles $\delta_\mrm{LOS}$
($z$-axis close to plane of sky) and $\delta_\mrm{jet}$ ($\theta_\mrm{jet}$
intrinsically large) the pre-merger jet will enclose only a small angle with
the LOS. This is consistent with the wings being almost perpendicular to the
post-merger jet and lying in the plane of sky, see Figs.~\ref{f_zrgs},
\ref{f_z-shapes}.

Within these limits we have three different possibilities to align both pairs
of jets and the LOS relative to each other, see \fgr{f_z-shapes} (angles are
positive in clockwise direction):
\begin{description}
\item[\textbf{a:}] $\delta_\mrm{LOS} < \delta_\mrm{jet}$ \enspace The LOS is
  closer to the merger plane than the pre-merger jet. The upper part of the
  jet is approaching us, the lower receding (dotted blue/dark-grey lobes).
\item[\textbf{b:}] $\delta_\mrm{LOS} > \delta_\mrm{jet}$ \enspace Same as
  \textbf{a}, but with the pre-merger jet closer to the merger plane than the
  LOS (dashed green/light-grey lobes).
\item[\textbf{c:}] $\delta_\mrm{LOS} > 0$, $\delta_\mrm{jet} < 0$ \enspace The
  receding part of the jet and the LOS are in the same hemisphere which is
  defined by the post-merger jet ($z$-axis) as polar axis (solid red/grey
  lobes). In this case it doesn't matter whether the LOS or the pre-merger
  jet is closer to the merger plane.
\end{description}
If we measure the lateral offset of the ridges of the wings we can deproject
the jets under assumptions for the bending radius. The thus obtained
expressions enable us to put some limits on this radius and to decide wich
orientation is most likely. This will be done in the following section.

\subsection{Deprojection of the Jets}
\label{s_depro}

With the observed projected angular extension in the sky of the straight part
of the \Z-shape, $\Theta_{\rm{p}}$, and the known distance $D$ of the galaxy
we can determine the angle $\theta_{\rm{jet}}$ between the pre- and
post-merger jets, provided we know the radius $r$ where the jet is bent.
\fgr{f_angle-a} shows the situation for orientation \textbf{a}
($\delta_{\rm{LOS}}<\delta_{\rm{jet}}$) of \fgr{f_z-shapes}, projected in a
plane perpendicular to the merger plane and containing both the LOS and
$z$-axis. Because $\varphi$ is assumed to be $0$, also the pre-merger jet lies
in this plane and the wings are perpendicular to the paper plane. As in
\fgr{f_z-shapes} the angles $\delta$ and $\theta$ in \fgr{f_angle-a} denote
the angle to the merger plane and $z$-axis respectively. See \fgr{f_angle-a}
for the meaning of the other quantities. Projecting the sum of the approaching
and receding jet $y_a$ and $y_r$ in the plane of sky yields the wanted
relation between $\Theta_{\rm{p}}$ = $\tan^{-1} y_\mrm{p}/D$, the bending
radius $r$ and the angles $\theta_\mrm{jet}$ and $\theta_\mrm{LOS}$.

\textbf{Orientation a:} (dotted blue/dark-grey lobes in \fgr{f_z-shapes}) With
 $a=D\tan\delta_\mrm{LOS}$ and the definition $k\equiv\tan\delta_{\rm{jet}} =
 h/r$ we get from \fgr{f_angle-a}
\[
\tan \delta_a = \frac{a-h}{D-r} = \frac{h-y_a}{r},
\]
which can be solved for $y_a$:
\beq
y_a = r\left(k - \frac{\tan\delta_\mrm{LOS} - k\,r/D}{1-r/D}\right).
\eeq
In the same way we get for the receding part of the jet
\beq
y_r = r\left(k - \frac{\tan\delta_\mrm{LOS} + k\,r/D}{1+r/D}\right),
\eeq
and hence for their sum along the $z$-axis in the limit that the distance to
the galaxy is much larger than the radius of the rotation stream
\beq
\begin{split}
y_z & =y_a+y_r=2r\left(
k-\frac{\tan\delta_{\rm{LOS}}-k\,(r/D)^2}{1-(r/D)^2}\right)\\
& \xrightarrow[D\gg r]{}\quad 2r(k-\tan\delta_{\rm{LOS}}).
\end{split}
\eeq Finally, after projecting $y_z$ into the plane of sky we obtain for the
projected lateral offset of the ridges \beq
y_{\rm{p}}=D\tan\Theta_\mrm{p}=y_z\cos\delta_{\rm{LOS}}=2r
\frac{\sin(\delta_{\rm{jet}} -\delta_{\rm{LOS}})}{\cos\delta_{\rm{jet}}}.
\label{eq_yp-a}
\eeq

Proceeding in the same way we obtain for the other possible orientations:

\textbf{Orientation b:} (dashed green/light-grey lobes)
Like \textbf{a}, with the difference of the angles $\delta_{\rm{jet}}$ and
$\delta_{\rm{LOS}}$ changing sign:
\begin{equation}
y_{\rm{p}}=2r \frac{\sin(\delta_{\rm{LOS}}
  -\delta_{\rm{jet}})}{\cos\delta_{\rm{jet}}}.
\label{eq_yp-b}
\end{equation}

\textbf{Orientation c:} (solid red/grey lobes) In notation of
\fgr{f_angle-a} like orientation \textbf{b} with the sign of
$\delta_{\rm{jet}}$ changed so that in the following expression the angle
varies in the range $0<\delta_\mrm{jet}<90^\circ$:
\begin{equation}
y_{\rm{p}}=2r \frac{\sin(\delta_{\rm{LOS}}
  +\delta_{\rm{jet}})}{\cos\delta_{\rm{jet}}}.
\label{eq_yp-c}
\end{equation}
As we pointed out in Sect.~\ref{s_orient} we expect $\delta_\mrm{LOS}$ to be
small because the post-merger jet is lying close to the plane of sky and
$\delta_\mrm{jet}$ is small because the intrinsic angle between both jets is
expected to be large. The latter is larger than zero though, because otherwise
we could not observe the offset of the wings.

\section{Application on NGC~326 and 3c~52}
\label{s_appben}

In this section we apply the results obtained above on the two observed
\Z-shaped XRGs to derive limits for the bending radius and to find the
possible orientations. Afterwards we will check whether the jet can still be
bent into the wings at the obtained distances.

\subsection{Bending radius and orientation}
\label{s_applic}

\begin{figure}
\ifpdf
  \includegraphics[width=84mm]{angles-ngc326.pdf}
\else
  \includegraphics[width=84mm]{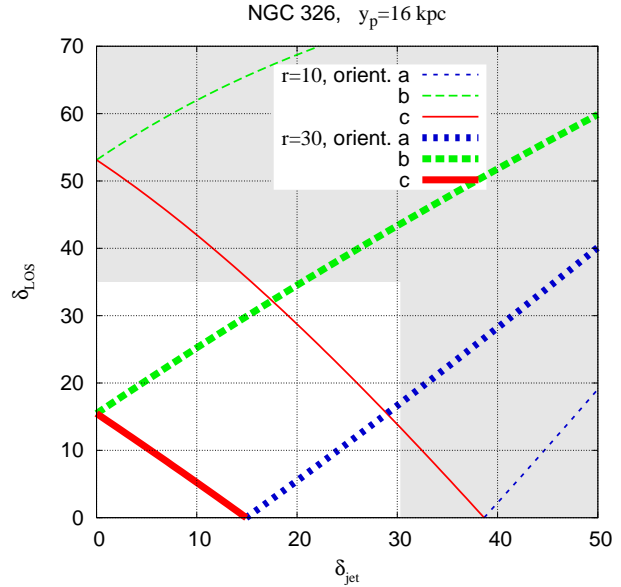}
\fi
\caption[]{For NGC~326 the angle of the LOS to the merger plane,
  $\delta_{\rm{LOS}}$, is plotted versus the angle $\delta_{\rm{jet}}$ between
  this plane and the secondary lobes, with the bending radius $r$ as
  parameter. Colors (grey-shades) of the lines refer to the different
  orientations. In the shaded area both or one angle becomes too large as to
  fulfil the condition of the primary lobes being close to the plane of sky
  and the angle between both pairs of lobes to be large.}
\label{f_res-ngc326} 
\end{figure}

Given the radius $r$ for the rotation field we can solve the above
Eqs.~(\ref{eq_yp-a}) to (\ref{eq_yp-c}) for $\delta_{\rm{LOS}}$ and plot it in
dependency of $\delta_{\rm{jet}}$ in order to find the most likely orientation
that minimizes both angles. The expressions we get are 
\begin{equation} 
\delta_\mrm{LOS} =
\begin{cases}
\phantom{-}\delta_\mrm{jet} -
\sin^{-1}\left(\frac{y_\mrm{p}}{2r}\cos\delta_\mrm{jet}\right), & 
\text{orientation\; \textbf{a}}\\
\phantom{-}\delta_\mrm{jet} +
\sin^{-1}\left(\frac{y_\mrm{p}}{2r}\cos\delta_\mrm{jet}\right), &
\text{\phantom{orientation\;} \textbf{b}}\\
-\delta_\mrm{jet} +
\sin^{-1}\left(\frac{y_\mrm{p}}{2r}\cos\delta_\mrm{jet}\right), &
\text{\phantom{orientation\;} \textbf{c}.}
\end{cases}
\label{eq_angles}
\end{equation}
The lateral offset $y_\mrm{p}$ of the ridges we take from observations of
NGC~326 and 3C~52.

\textbf{NGC~326:} The distance to this source is about
$160\,\mrm{Mpc}$. Therefore the projected angular size of the middle part of
the `\Z', $\Theta_\mrm{p} = 20''$, translates into a projected length of about
$y_\mrm{p} = 16\,\mrm{kpc}$. \citet{schiminovich94} and \citet{charmandaris00}
have detected dense clouds that contain both \HI and molecular gas in
Cen~A. With an assumed distance of $\sim 3.5\,\mrm{Mpc}$ to Cen~A they locate
the gas at a radius of about $10\,\mrm{kpc}$ to the center.  If we use this as
the radius $r$ of the rotation field and plot $\delta_\mrm{LOS}$ in dependency
of $\delta_\mrm{jet}$ (\eqn{eq_angles}) we get the thin curves in
\fgr{f_res-ngc326}. Again the dotted blue/dark-grey line represents the
solutions for orientation \textbf{a}, the dashed green/light-grey line for
\textbf{b} and the solid red/grey line for \textbf{c}. Because the intrinsic
angle is expected to be larger than $\sim60^\circ$, $\delta_\mrm{jet}$ is less
than $30^\circ$ (left from the shaded area). This limit excludes orientation
\textbf{a} which has a minimum of about $\delta_\mrm{jet}\approx
38^\circ$. The smallest pair of angles in configuration \textbf{b} is
$0^\circ$ for $\delta_\mrm{jet}$ and $\delta_\mrm{LOS}\approx 53^\circ$. Such
a large angle violates the condition that primary lobes have to be close to
the plane of sky in order to detect an \X-shape (region below the shaded
area). Hence also this orientation is ruled out and the only remaining
possibility is \textbf{c}, where the LOS and the approaching part of the jet
are in different hemispheres relative to the post-merger jet. To minimize both
angles we get $\delta_\mrm{LOS} = \delta_\mrm{jet} \approx 23.6^\circ$, with
not too much range left for them on curve \textbf{c}, if we assume the shaded
areas outside the inner rectangle as not permitted. Being pushed to the limit
for a radius of $10\,\mrm{kpc}$, the situation is much less restrictive if we
allow for a larger radius. The thick curves in \fgr{f_res-ngc326} show the
results if the jet is bent in a distance of $r=30\,\mrm{kpc}$. On branch
\textbf{c} both angles are much smaller, with an upper limit of about
$15^\circ$ for $\delta_\mrm{jet}$. For the larger radius also the other
orientations \textbf{a} and \textbf{b} are possible. These two branches appear
in the allowed, not-shaded region of \fgr{f_res-ngc326} only for $r\gtrsim
14\,\mrm{kpc}$.

\begin{figure}
\ifpdf
  \includegraphics[width=84mm]{angles-3c52.pdf}
\else
  \includegraphics[width=84mm]{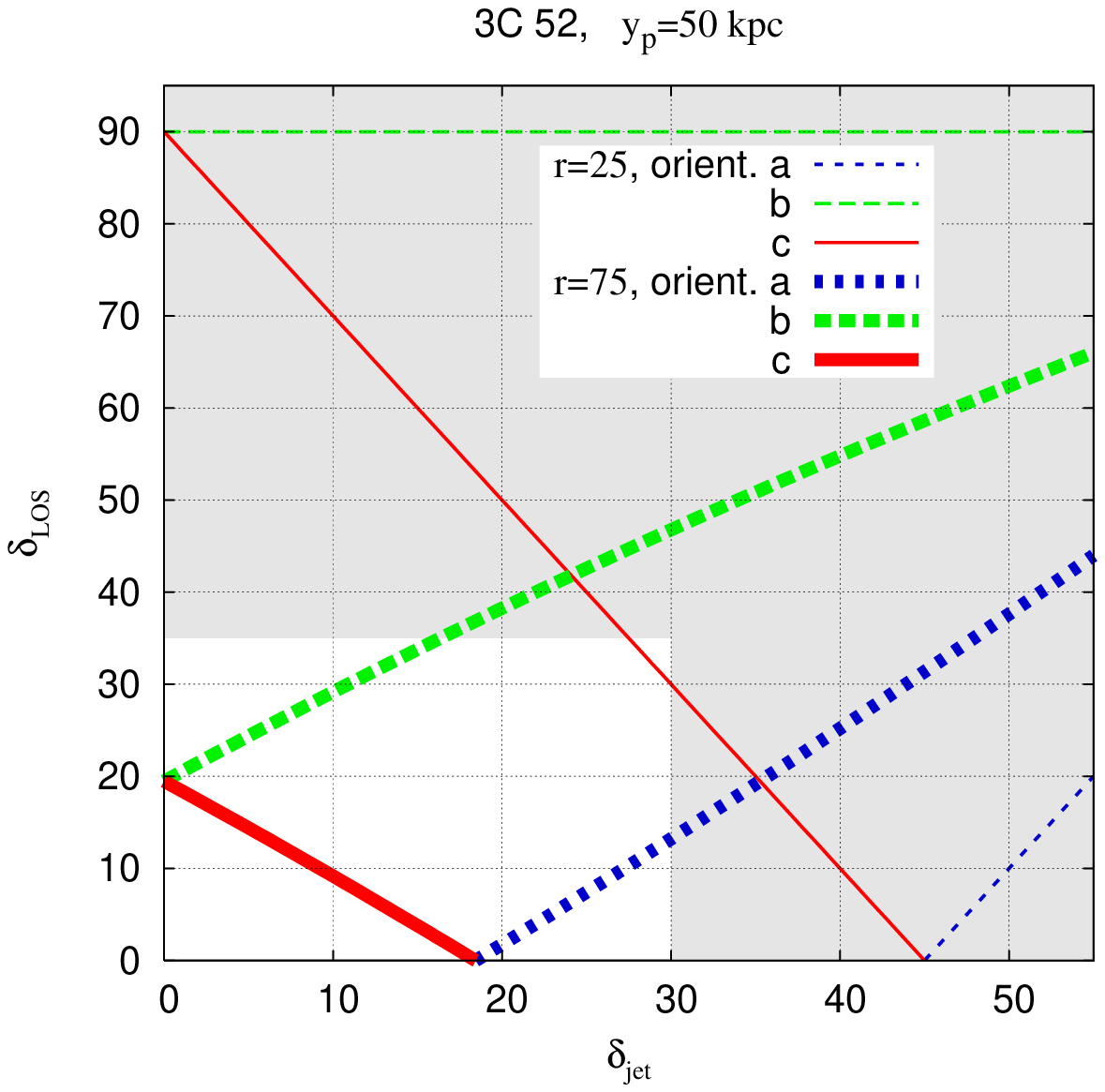}
\fi
\caption[]{The same as \fgr{f_res-ngc326}, but for 3C~52.}
\label{f_res-3c52} 
\end{figure}

\textbf{3C~52:} For 3C~52, in a distance of about $1\,\mrm{Gpc}$, with
$\Theta_\mrm{p} = 10''$ the projected offset of the wings is $y_{\rm{p}} =
50\,\mrm{kpc}$. This is five times the length of the assumed radius of
$10\,\mrm{kpc}$. But for \eqn{eq_angles} to have a solution, the argument of
$\sin^{-1}$ has to be less than $1$ ($\delta_\mrm{jet}$ varies in a range
where the cosine is positive), leaving us with the condition
\[
\cos\delta_\mrm{jet} \le \frac{2r}{y_\mrm{p}}.
\]
For $r=10\,\mrm{kpc}$ this is fulfilled only if $\delta_\mrm{jet}$ is larger
than $66^\circ$, in contradiction with the intrinsic angle between the
jet-pairs to be large. So for this small radius none of the orientations is
within the allowed rectangle of \fgr{f_res-3c52}. Only for $r$ larger than
$y_\mrm{p}/2 = 25\,\mrm{kpc}$ the full range between $0^\circ$ and $90^\circ$
becomes mathematically possible for $\delta_\mrm{jet}$. For this radius the
matching angles of the LOS are too large and none of the orientations provides
a satisfying solution, see the thin lines in \fgr{f_res-3c52}. A larger
bending radius can solve the problem again. If we gradually increase the
radius, first branch \textbf{c} offers physically reasonable results for both
angles ($r=24\,\mrm{kpc}$) before the other two branches enter the allowed
region for $r>44\,\mrm{kpc}$. The thick lines in \fgr{f_res-3c52} show the
results for $r=75\,\mrm{kpc}$, with all orientations being possible.

These results show that the bending of the lobes into the wings occurs already
much earlier during the merger in distances of more than $20$ to maybe even
$100\,\mrm{kpc}$. The smaller the radius is, the more likely the jets have
orientation \textbf{c} relative to us, i.e. the receding pre-merger jet and
the LOS are in the same hemisphere that is defined by the post-merger jet
(axis of orbital angular momentum) as polar axis.

\subsection{Bending the jet}
\label{s_bending}
If the bending of the jet happens at the distances suggested above, we have to
verify that the rotating stream is able to exert a large enough pressure on
the jet at these radii to bend it into \Z-shape as it has been shown by
\citetalias{gopal-krishna03} for a radius of $10\,\mrm{kpc}$. The bending of
the jet is described by Euler's equation, which for a steady state flow reads
\begin{equation}
\rho_\mrm{jet} (\mbf{v}_\mrm{jet}\bvec{\nabla})\mbf{v}_\mrm{jet}  =
-\bvec{\nabla} P_\mrm{ISM},
\label{eq_euler}
\end{equation}
with the gradient of the ram pressure of the ISM being applied transverse to
the beam \citep{odonoghue93}. The velocity of the jet is assumed to change by
order of itself over the bending scale $l_\mrm{bend}$, so that the left hand
side of the equation can be written as $\rho_\mrm{jet} v_\mrm{jet}^2 /
l_\mrm{bend}$. As \citeauthor{odonoghue93} point out a pressure gradient that
provides a centripetal acceleration $v_\mrm{jet}^2/l_\mrm{bend}$ around a
curve with radius $l_\mrm{bend}$ gives the same result for the left hand side
of \eqn{eq_euler}. The pressure gradient due to ram pressure on the right hand
side, $\bvec{\nabla} P_\mrm{ISM}$, can be approximated by $\rho_\mrm{ISM}
v_\mrm{ISM}^2/l_\mrm{press}$ if $v_\mrm{ISM}$ is the relative velocity between
the ISM and the galaxy, i.e. the rotation velocity. $l_\mrm{press}$ is the
length scale over which the ISM exerts the ram pressure on the jet and is
taken to be the radius of the jet ($R_\mrm{jet}$) at the bending point
($r$). Hence we can rewrite Euler's equation in the approximation for jet
flows as
\begin{equation}
\rho_\mrm{jet}\frac{v_\mrm{jet}^2}{l_\mrm{bend}} =
\rho_\mrm{ISM}\frac{v_\mrm{ISM}^2}{l_\mrm{press}}.
\label{eq_euler-jet}
\end{equation}
Assuming that the clouds in Cen~A represent the properties of the ISM
reasonably well the following reference values can be used: $n_\mrm{ISM} =
\rho_\mrm{ISM}/(1.4\,m_\mrm{p}) = 0.1\,\mrm{cm}^{-3}$ and $v_\mrm{ISM} =
100\,\mrm{km/s}$, with $l_\mrm{press} = R_\mrm{jet} \sim 1\,\mrm{kpc}$. The
jet is assumed to be semirelativistic ($v_\mrm{jet}=10^5\,\mrm{km/s}$) and
made of ordinary proton-electron plasma that is bent over a scale of
$l_\mrm{bend} = 10\,\mrm{kpc}$. Thus for the density of a jet that is bent
into \Z-shape by ram pressure a density of $n_\mrm{jet} =
10^{-6}\,\mrm{cm}^{-3}$ is obtained.

Now we have to scale up the values obtained at $r_1 = 10\,\mrm{kpc}$ to larger
distances $r_2$ such that \eqn{eq_euler-jet} is still fulfilled. In the
following the additional indices $1$ and $2$ refer to the quantities in
distance $r_1$ and $r_2$ respectively. If the half-opening angle of the jet is
$\vartheta$, then its spherical surface perpendicular to the direction of
propagation in a distance $r$ is $A=2\pi r^2 (1-\cos\vartheta)$. Taking the
flux of momentum along the jet $\rho_\mrm{jet} v_\mrm{jet}^2 A$ to be constant
we find
\begin{equation}
\rho_{\mrm{jet}_2} v_{\mrm{jet}_2}^2 = \rho_{\mrm{jet}_1}
v_{\mrm{jet}_1}^2 \left(\frac{r_1}{r_2}\right)^2 
\label{eq_flux}
\end{equation}
With this expression, \eqn{eq_euler-jet}, the relation $R_\mrm{jet_2} =
R_\mrm{jet_1} r_2/r_1$ and the assumption that $l_\mrm{bend}$ scales linearly
with $r$ we finally obtain for the density at $r_2$
\begin{equation}
\rho_\mrm{ISM_2}=\rho_\mrm{ISM_1} \left(\frac{r_1}{r_2}\right)^2
\left(\frac{v_\mrm{ISM_1}}{v_\mrm{ISM_2}}\right)^2.
\label{eq_scale}
\end{equation}
Hence, to be able to bend the jet in, say $r_2 = 50\,\mrm{kpc}$, the density
of the ISM can be 25 times less than at $r_1= 10\,\mrm{kpc}$. For an annulus
with a width from $50$ to $60\,{\rm kpc}$ and of $10\,{\rm kpc}$ height that
density corresponds to a total mass of $\sim3.4\times 10^9 M_{\odot}$. For a
velocity of about $200\,\mrm{km/s}$, as has been observed in such distances
(see Sect. \ref{s_discussion}) the mass is reduced by another factor of four,
so that it is less than $10^9\,M_\odot$. At $r_1$ the mass in a ring with
inner and outer radius $7.5$ and $12.5\,\mrm{kpc}$ respectively and a hight of
$5\,\mrm{kpc}$ is $\sim4\times 10^9 M_{\odot}$. Thus about the same mass that
is required in a rotating stream with radius $10\,\mrm{kpc}$ to bend the jet
into a \Z-shape is sufficient to bend the jet at much larger radii. The
required mass and velocity at such distances is in agreement with observations
(see next section). Hence our results show that the ram pressure of the
rotating gas in a distance of $\sim 50\,\mrm{kpc}$ is indeed strong enough to
bend the jet in \Z-shape, as is required by the geometrical arguments above.

\section{Discussion}
\label{s_discussion}
\subsection{Geometry and dynamics of the jet}
\label{s_discussion1}
In the previous sections we showed that the bending of the pre-merger jet into
\Z-shape, as proposed by \citetalias{gopal-krishna03} within the merger model,
must happen on distances larger than the $10\,\mrm{kpc}$ that they have
suggested. Because the half-thickness of the rotating gas stream will not be
larger than its radius, it has to pass through the polar regions of the
primary galaxy in order to bend the jet and thus pre- and post-merger jets are
approximately perpendicular to each other. This corresponds to the maximum of
the distribution of the intrinsic angle ($\theta_\mrm{jet} = 90^\circ$)
between both pairs of lobes in XRGs, if the directions of their propagation
are uncorrelated. But this is exactly what we expect in the merger model for
XRGs, if the spin of the post-merger \BH is dominated by the orbital angular
momentum of the binary, as pointed out in Sect.~\ref{s_orient}. In ZRGs a
larger bending radius is in favour of a larger angle $\theta_\mrm{jet}$
between the jets. For example $y_\mrm{p}=50\,\mrm{kpc}$ has been observed in
3C~52. Trying to minimize the the half-height $h$ for $r=10\,\mrm{kpc}$,
i.e. maximizing $\theta_\mrm{jet}$, we obtain with orientation \textbf{a}
$h=y_\mrm{p}/2=2.5 r$ at $\theta_\mrm{jet}=21.8^\circ$ and
$\delta_\mrm{LOS}=0$. In case \textbf{b} there is no solution at all for
$y_\mrm{p}\ge 2r$ and for \textbf{c} we get $h=2.3 r$ at $\theta_\mrm{jet} =
25.6^\circ$ and $\delta_\mrm{LOS} = 23.6^\circ$. This is in direct
contradiction with a slim gas stream and with the assumption of a large angle
between both jet pairs and could be solved with a larger bending radius
(Sect.~\ref{s_applic}).

As \citetalias{gopal-krishna03} estimated, the ram pressure of the rotation
field at $10\,\mrm{kpc}$ radius is strong enough to bend a jet with power
close to the FR~I/II transition into \Z-symmetry. A stronger jet would not be
much deflected by the rotating gas-stream. Because the wings are in the plane
of sky and extend almost perpendicular from the post-merger jet, the
pre-merger jet would also have to be close to the plane of sky and hence
distinguishable from the primary jet, as we pointed out at the end of
Sect. \ref{s_orient}. However, this has not been observed and thus also the
appearance and morphology of the source argue for weaker jets which can be
deflected by a large angle. But the jet should not be too weak as well,
because then it would be just dragged along with the circular gas stream and
hence have a maximum projected extension of the bending radius. Depending on
the angle of the LOS to the merger plane we might see the curvatures of the
wings following the circular motion and exhibiting inversion symmetry relative
to the center as depicted in \fgr{f_z-shapes} (of course jets deflected at
smaller angles also show inversion symmetry, but they can more easily deviate
from that due to interaction with the ambient medium). In case of NGC~326 and
3C~52 this means a bending radius of at least $50$ and $100\,\mrm{kpc}$
respectively. It is more likely that a weak jet would be bent at smaller radii
by the inspiralling gas stream, where the jet has a larger power and hence
would probably be deflected but not dragged along with the rotational
motion. In Sect. \ref{s_bending} we showed that a gas stream with a similar
mass content as that required for bending at $10\,\mrm{kpc}$ is able to bend
the same jet at larger distances of about $50\,\mrm{kpc}$. After spiralling
further inside to smaller radii the secondary galaxy will have suffered more
stripping of material and the gas stream becomes weaker while the jet becomes
stronger. The requirements for the stream at $50\,\mrm{kpc}$ are in good
agreement with observations (Sect.~\ref{s_discussion2}). Hence the conclusion
by \citetalias{gopal-krishna03} that the bending of jets into \Z-shape happens
at a jet-power close to the FR~I/II transition holds also at larger radii.

\subsection{Evidence for the required streams}
\label{s_discussion2}
In the present model we assume XRGs and ZRGs to be merger products. As such it
is expected that the secondary galaxy, while spiralling inwards, induces a
stream of gas and dust on scales of tens of $\mrm{kpc}$ in the primary
galaxy. This stream is due to matter of the primary galaxy dragged along by
the secondary as well as matter stripped off from the secondary galaxy. Now
looking for such streams in other sources shows that they have been observed
in various objects. These streams are always related to a merger between two
galaxies. This is also in very good agreement with numerical simulations of
mergers which produce tidal tails and streams with the properties required in
our model ($\rho, v, r$) and hence lends strong support to it. In the
following we compile some information of these sources. We use
$H_0=70\,\mrm{km/s\,Mpc}$ and scaled the values from the cited papers
accordingly. A summary is given in Table~\ref{t_only}.

Recent \HI observations of M~31 by \citet{thilker04} show a circumgalactic
cloud population in $\sim50\,\mrm{kpc}$ distance. These clouds are moving with
a velocity component along the LOS of
$v_\mrm{syst}\phantom{}^{+128}_{-215}\,\mrm{km/s}$, matching the velocity
extent of the disk of M~31. Though the \HI content of the halo cloud
population is estimated to be only $\sim3\negmedspace-\negmedspace7\times
10^{7}\,M_\odot$, it might trace more substantial amounts of ionized gas and
dark matter. As an obvious source of the high-velocity \HI gas the authors
give tidal stripping from mergers in agreement with \citet{brown03}, who
relate the young halo to a major merger or several minor mergers.

\citet{braun03} conducted a \HI survey and found significant positional
offsets exceeding $10\,\mrm{kpc}$ in some of the sources, what they attribute
to tidal interaction. While the mean observed offset of \HI is about
$66\,\mrm{kpc}$, in NGC~1161 \HI is observed in $110\,\mrm{kpc}$ distance to
the center at speeds that differ by $\sim200\,\mrm{km/s}$ from the systemic
velocity. The \HI mass is estimated to be about $1.8\times 10^9\,M_\odot$.

The \HI detected by \citet{vangorkom86} in NGC~1052, an active elliptical
galaxy, is distributed in a disk that extends
$20\negmedspace-\negmedspace25\,\mrm{kpc}$ along the minor axis and is seen
almost edge on. The gas has a circular velocity of $\sim200\,\mrm{km/s}$ that
is roughly constant with radius. The \HI mass is about
$5.7\times10^8\,M_\odot$ and shows an outer structure that resembles tidal
tails, which \citeauthor{vangorkom86} attribute to a merger about
$10^9\,\mrm{yr}$ ago.

In another early type galaxy, IC~5063 which has a Seyfert~2 nucleus,
\citet{morganti98} detect \HI that to first order is distributed in disk of
about $28\,\mrm{kpc}$ radius. This disk is oriented very similar to a system
of dust lanes and in projection rotates at $240\,\mrm{km/s}$ with an \HI mass
of $4.2\times 10^9\,M_\odot$. Optical data from previous observations
\citep[e.g.,][]{danziger81} revealed ionized gas that also lies in a disk,
which is extending to $\sim 14.4\,\mrm{kpc}$. The faint structures in the
outer regions could be tidal arms and the origin of \HI is most likely a
merger between spiral galaxies as in the other sources.

\begin{table}
\caption[]{Observed masses and properties of rotating gas streams.}
\label{t_only}
\begin{tabular}{@{}lrrrr}

\hline

Source & \multicolumn{1}{c}{$\log(\frac{M_\textrm{\HI}}{M_\odot})$} &
\multicolumn{1}{c}{$r$} & \multicolumn{1}{c}{$v$} & \multicolumn{1}{c}{$D$} \\

& & \multicolumn{1}{c}{(kpc)} & \multicolumn{1}{c}{(km/s)} &
  \multicolumn{1}{c}{(Mpc)} \\

\hline 

M~31 & $7.5\negmedspace-\negmedspace7.9^{(a)}$ & $50$ &
$\phantom{}^{+128}_{-215}$ & 0.77\\

NGC~1161 & $9.3^{\phantom{(c)}}$ & $110$ & $200$ & $28$ \\

NGC~1052 & $8.8{\phantom{(c)}}$ & $20\negmedspace-\negmedspace25$ & $200$ &
$21$ \\

IC~5063 & $9.6{\phantom{(c)}}$ & $28$ & $240$ & $48.6$ \\ 

NGC~3108 & $9.4{\phantom{(c)}}$ & $30$ & $290$ & $38$ \\ 

Cen~A & $8.2{\phantom{(c)}}$ & $34$ & $250$ & $8$ \\

NGC~5266 & $10.1^{(b)}$ & $51\negmedspace-\negmedspace100$ & $270$ & $44$ \\

NGC~4650A & $10.1^{(c)}$ & $10$ & $120$ & $41$ \\

IC~1182 & $10.3{\phantom{(c)}}$ & $60$ & $100$ & $146$ \\

Arp~105 & $9.8{\phantom{(c)}}$ & $100$ & $200$ & $125$ \\

NGC~7252 & $9.6{\phantom{(c)}}$ & $60$ & $100$ & $67$ 
\end{tabular}

\medskip

$H_0=70\,\mrm{km/s\,Mpc}$ has been used throughout. $M_\textrm{\HI}$ is the
\HI mass in disks or streams. This gas has been observed in distance $r$ to
the center and is moving at circular velocities $v$. Only for M~31 the
velocity has not been deprojected. $(a)$ Mass of the halo cloud population,
possibly tracing more substantial amounts of ionized gas and dark
matter. $(b)$ Total \HI mass. $(c)$ Luminous mass in polar structure. For
details and references see text.

\end{table}

Among the five elliptical galaxies that \citet{oosterloo02} observed they
detected $2.3\times 10^9\,M_\odot$ of \HI in NGC~3108 that is distributed in a
disk-like structure perpendicular to the optical major axis of the galaxy and
extends to $\sim 30\,\mrm{kpc}$. They assume an inclination of $70^\circ$ and
thus obtain for the rotation velocity $290\,\mrm{km/s}$, which appears to be
constant from $\sim1\,\mrm{kpc}$ to the very outer regions. Within the central
$1'$ the disk seems to have a hole that is filled by a disk seen in emission
from ionized gas. While the boxy outer isophotes indicate that also NGC~3108
has undergone a major merger, the regular and settled appearance of the disk
suggest that it happened some $10^9\,\mrm{yr}$ ago.

Cen~A is a giant elliptical galaxy with an active nucleus that shows strong
radio lobes on both sides of a dust lane which is aligned with the minor axis
\citep{clarke92} and a warped gaseous disk which is seen in optical and \HI
emission \citep{dufour79,vangorkom90}. \citet{schiminovich94} find the \HI
morphology to be closely correlated with diffuse shells seen in the optical
range \citep{malin83} and estimate the total mass in the shells to be $\sim
1.5\times 10^8\,M_{\odot}$. The position-velocity (PV) plot of \HI in the
shells is well fitted by a single ring with uniform rotation velocity ($\sim
250\,{\rm km/s}$) and the rotation axis being roughly perpendicular to that of
the inner \HI disk. The rotation curve is flat out to $15'$, what corresponds
to a radius of $34\,\mrm{kpc}$ for $D=8\,\mrm{Mpc}$, using a redshift of
$z=0.001825$ (note that \citeauthor{schiminovich94} used $D=3.5\,\mrm{Mpc}$,
and hence $r \sim 10\,\mrm{kpc}$, what we initially used as bending radius in
Sect.~\ref{s_applic}). As possible explanation for the misalignment bewteen
the rotation axis of the \HI in the shells and the disk they suggest a merger
which is not proceeding in the plane of Cen~A, and differential precession of
the stripped material.  Later \citet{charmandaris00} suggested that the
morphology of the shells is a combination of both, phase wrapping of tidal
debris on nearly radial orbits \citep{quinn84} and spatial wrapping of matter
in thin disks for mergers with large angular momentum
\citep{dupraz87,hernquist89}. They also detect CO emission in the shells and
associate it with the \HI gas which shows the same velocity signatures and
deduce a $\text{H}_2$-mass of $4.3\times 10^7\,M_\odot$.

In NGC~5266, a bright E4 galaxy, \citet{morganti97} find \HI gas distributed
in two perpendicular disks. The inner disk is aligned with the dust lane and
fills the hole of $\sim2'$ diameter of the outer disk, which extends to $4'$
($51\,\mrm{kpc}$). For the rotation velocity of this disk they obtain
$270\,\mrm{km/s}$, which is constant in radius, and for the total \HI mass
$1.2\times 10^{10}\,M_\odot$. They point out that the \HI distribution is
similar to that observed in Cen~A, with most of \HI being associated with the
dust lane but a different kinematical behaviour at larger distances and
forming a ring that is roughly perpendicular to the dust lane. This is unlike
than in most polar-ring and dust-lane galaxies. The outer parts of \HI,
extending to $\sim100\,\mrm{kpc}$, could be a settled ring but are rather
interpreted as tidal tails that formed during a merger of two gas-rich spiral
galaxies. Numerical simulations by \citet{hibbard95} (see later in this
section) have shown that after gas piling up in the center and fuelling a star
burst, the fraction that remains at larger distances in tidal tails will
settle in a disk or a ring, depending on the initial conditions. Earlier
observations in the optical range indicate that the kinematic axes of stars
and gas are orthogonal, with the gas in the dust lane rotating about the
optical major axis \citep{caldwell84}. \citet{goudfrooij94} could show that
ionized gas lies in a ring that is clearly associated with the dust ring
detected. In CO observations \citet{sage93} found that the molecular gas is
also distributed in a ring that is co-rotating with the ionized gas at
velocities of $270\negmedspace-\negmedspace300\,\mrm{km/s}$ within $1'$
($\sim13\,\mrm{kpc}$) and has a mass in $\text{H}_2$ of
$2.7\times10^9\,M_\odot$.

Recently \citet{swaters03} observed the polar ring galaxy NGC~4650A and found
the velocities of both, stars and gas in the polar ring component to be
closely correlated. The ring is seen close to edge on and rotates close to its
outer parts ($\sim 10\,\mrm{kpc}$) with a velocity of $120\,\mrm{km/s}$. The
flatness of the rotation curve suggests that the gas and stars are rather
distributed in a disk than a norrow ring. This is supported by the results of
\citet{bekki98} who simulated a dissipational polar merger of two disk
galaxies of about the same mass. In his numerical experiments he could
reproduce polar ring galaxies, with the intruding galaxy being transformed
into a S0-like host and the victim into a narrow polar ring. As standard model
\citeauthor{bekki98} used $10^{10}\,M_\odot$ and $10\,\mrm{kpc}$ for the disk
mass and radius respectively. These are quite small values and we scaled the
model to masses in the range of $10^{11-12}\,M_\odot$. Then crude estimates of
the size, velocity and mass of the polar ring result in the ranges that are
required by our model, i.e. $30\negmedspace-\negmedspace100\,\mrm{kpc}$,
$100\negmedspace-\negmedspace300\,\mrm{km/s}$ and $\sim 1/100\,M_\mrm{disk}$
respectively. \citeauthor{bekki98} notes that he seemed to have failed to
reproduce annular polar ring galaxies like NGC~4650A, unless the annular ring
component with an apparent hole in the center is a part from the galactic
disk, in good agreement with the conjecture by \citet{swaters03} based on
their observations.

In other numerical simulations of mergers \citet{bournaud04} compared their
results with the kinematics of tidal tails in interacting galaxies. Their main
goal is to distinguish whether apparent massive condensations close to the
tips of the tails are real or caused by projection effects. The PV plots can
qualitatively distinguish both possibilities: Either the difference between
the systemic velocity and the tidal tail increases with position along the
tail, reaching a maximum at its end, or it passes through a maximum before it
decreases and even turns back to closer positions at smaller speeds, thus
following a loop. We are more interested in the latter case which can give us
information about the azimuthal component of the velocity in the tail, that we
assume to be edge on. The matter in the tail is not streaming along its
spatial extension and also has a velocity component perpendicular to the
tangential one. As the velocity component aligned with the LOS is measured
along the tail with increasing distance to the center, the velocity
increases. Before the tip is reached the velocity assumes a maximum when the
velocity in the tail is aligned with the LOS. As the observer moves on to the
tip where only the azimuthal component is aligned with the LOS, the velocity
decreases. Comparison with observations show that IC~1182 and the northern
tail of Arp~105 fall in this category with a circular velocity of $\sim
100\,\mrm{km/s}$ in about $60\,\mrm{kpc}$ distance from the center and $\sim
200\,\mrm{km/s}$ at $\sim 100\,\mrm{kpc}$ respectively. The corresponding \HI
mass at the end of the tails they estimate to be about
$1.8\times10^{10}\,M_\odot$ and $6.5\times10^{9}\,M_\odot$ respectively.

NGC~7252 is a late stage merger of two gas-rich disk galaxies
\citep[e.g.,][]{dupraz90,wang92,hibbard94}. The morphology and kinematical
properties of the tidal tails have been tried to be reproduced with numerical
simulations by \citet{hibbard95}. The PV plot in a previous paper by
\citet{hibbard94} shows that the tails have a maximum velocity along the LOS
of $100\,\mrm{km/s}$ at $\sim 60\,\mrm{kpc}$ before turning back in a loop in
the PV-plane. The \HI mass in the tails is estimated to
$\sim4\times10^9\,M_\odot$. The best-fit model of \citet{hibbard95} succeeds
in reproducing both the observed spatial morphology and the velocity structure
of the tails. By the time the simulated merger fits best the observations
($\sim8.3\times10^8\,\mrm{yr}$) about $4\times10^9\,M_\odot$ \HI is still in
the tails, half of which will fall back to radii $\lesssim20\,\mrm{kpc}$
during the next $4.3\times10^9\,\mrm{yr}$. We used the angular momentum --
radius plot to compute the circular velocity of the matter falling back to
$35\,\mrm{kpc}$ from the range between $65$ and $130\,\mrm{kpc}$ and obtain a
velocity range from $100$ to $200\,\mrm{km/s}$ respectively. Hence for a long
time a stream of gas strong enough to bend the jet is maintained in a distance
of about $35\,\mrm{kpc}$. Some of the falling back material will form loops or
shells and other structures.  In more recent simulations \citet{mihos98}
examined the effect of dark matter halo potentials on the morphology and
kinematics of tidal tails. They basically confirm the previous results from
\citet{hibbard95} which are relevant for our purposes. From the mass of the
initial \HI disk $10\negmedspace-\negmedspace20\%$ and about
$15\negmedspace-\negmedspace20\%$ of the initial total disk mass ends as
stellar mass in the tidal tails. Scaled to the Milky~Way this is about
$6.6\times10^9$ and $7.2\times10^9\,M_\odot$ respectively that circulates at
velocities of $220\,\mrm{km/s}$ at distances more than $100\,\mrm{kpc}$ around
the center.

Both, the observations of galaxies showing signatures of mergers about
$10^9\,\mrm{yr}$ ago, and simulations of merging galaxies are in good
agreement with our model. With the properties shown in Table~\ref{t_only} the
merger induced gas stream is able to bend a jet with power close to the
FR~I/II transition into \Z-shape on distances larger than $30\,\mrm{kpc}$.

\section{Conclusions}
\label{s_final}

In the present article we are investigating in the possible orientation and
geometry of \Z-shaped radio galaxies (ZRGs). The formation of objects of this
class has been explained by \citetalias{gopal-krishna03} within the framework
of a merger model. As the secondary galaxy spirals in towards the common
center of mass it generates a rotational velocity field of gas and dust in its
wake that is made up by matter stripped off from the secondary galaxy and
matter dragged along from the ambient medium of the primary galaxy. If the
trajectory passes through the polar region of the primary galaxy, its jet can
be bent into a \Z-symmetric shape, depending on the relative pressure between
the gas stream and the jet. Thus for these sources the spin of the primary \BH
and the orbital angular momentum $\mbf{L}_\mrm{orb}$ of the binary are
necesserily roughly perpendicular to each other. Provided that the spin of the
post-merger \BH is dominated by $\mbf{L}_\mrm{orb}$ and that jets are aligned
with the spin of the \BH at their base, consequently the pre- and post-merger
jets will also be perpendicular to each other. While this should be true for
every ZRG, this holds for XRGs only on average since the direction of the
$\mbf{L}_\mrm{orb}$ and the pre-merger spin of the \BH are uncorrelated, as
explained in Sect.~\ref{s_orient} \citep{rottmann01,zier01,zier02}. We used
this argument for large angles between the jet pairs and the assumption that
the post-merger lobes are close to the plane of sky, because they are
similarly luminous, to deproject the jets with respect to the LOS. Applied to
the ZRGs NGC~326 and 3C~52 our results showed that to fulfil both conditions
the bending of the jet must happen on scales between about $30$ and
$100\,\mrm{kpc}$. One important result is that the possibility to see the
source at one of the possible three orientations, indicated in
\fgr{f_z-shapes}, depends on the bending radius (Sect.~\ref{s_applic}). To
maintain a large angle between the jet pairs and a primary jet close to the
plain of sky we used the following limits for the angles: $\theta_\mrm{jet} =
90^\circ -\delta_\mrm{jet} > 60^\circ$ and $\delta_\mrm{LOS} < 35^\circ$,
which we think to be quite conservative. For very small bending radii $r$ no
solution is in the allowed range (white rectangle in Figs.~\ref{f_res-ngc326}
and \ref{f_res-3c52}). As we increase $r$ first orienation \textbf{c}, where
the LOS and the approaching part of the pre-merger jet are in different
hemispheres that are defined by the post-merger jet as polar axis, appears in
the allowed region for $r \gtrsim y_\mrm{p}/2$ (i.e. $8$ and $25\,\mrm{kpc}$
for NGC~326 and 3C~52 respectively). If we further increase $r$ both the other
orientations also appear in the allowed rectangle at roughly the same radius
$r\gtrsim y_\mrm{p}\sqrt{3}/2$ ($14$ and $44\,\mrm{kpc}$ for NGC~326 and 3C~52
respectively). Knowing the correct orientation we also know the sense of
rotation, i.e. $\mbf{L}_\mrm{orb}$, and consequently the direction of the spin
of the post-merger \BH \citep{chirvasa02,zier02,biermann02}. In future work
this could be compared with the spin inferred from circular polarization
measurements at cm-wavelengths, as has been dicussed by \citet{ensslin03} and
suggested by \citetalias{gopal-krishna03}.

The radius of the bending will depend on the relative pressure between the jet
and the gas stream. In Sect.~\ref{s_bending} we showed that for a jet with
a power close to the FR~I/II transition this happens on scales of
$50\,\mrm{kpc}$, in agreement with the results from the geometrical arguments
above. Thus the conclusion by \citetalias{gopal-krishna03} that the jet is
bent at a power close to the FR~I/II transition is also valid at radii in a
range of $30$ to $100\,\mrm{kpc}$. In fact in Sect.~\ref{s_discussion1} we
showed that neither very strong nor weak jets are in agreement with the
geometry of ZRGs.

ZRGs can not be explained by the rapid jet reorientation from instabilities in
an accretion disk, what is also considered as one possible formation mechanism
of XRGs \citep{dennett-thorpe02}, and might not be easily reconciled with the
observed distribution of angles between the jet pairs. Since ZRGs are a subset
of XRGs their existence strongly supports the merger model in favour of the
accretion model as formation mechanism of XRGs. In this context our result
that the angles between both jet pairs have to be large in ZRGs and are large
on average in XRGs, as has been observed \citep{rottmann01}, further
strengthen the merger model.

This in turn will have some impact on the ``final parsec problem'', i.e. the
conjecture that after a merger of two galaxies the merging of the \BH{}s
stalls in a distance of about $0.01$ to $1\,\mrm{pc}$ \citep{begelman80}. At
this distance dynamical friction is inefficient and gravitational radiation
still unimportant so that slingshot ejection of individual stars provides the
only mechanism to extract energy and angular momentum from the binary
\citep{zier01}. If there are no stars with small enough pericenters as to
interact with the binary (i.e. loss-cone depletion) the \BH{}s are prevented
from further merging. But contrary to that the existence of XRGs and ZRGs
shows that the binary has merged. In ZRGs they probably merge on timescales of
some $10^8\,\mrm{yr}$ after the bending of the jet in a distance of
$50\,\mrm{kpc}$. While in XRGs the binary could have stalled for a long time
on scales of $1\,\mrm{pc}$, in ZRGs the merger must have been completed after
the bending in a time short enough to maintain the rotational gas stream and
that we are still able to see the bended lobes, which are fading away and
undergo spectral ageing \citep{rottmann01}. Thus, in a way, the bending starts
a stop watch for the rest of the merger.


\section*{Acknowledgements}
I would like to thank Gopal-Krishna for helpful and valuable discussions on
the rotational motion of the ISM.

I am happy to have the opportunity to thank the RRI for the generous support
and very kind hospitality.

I also would like to thank Wolfram Kr\"ulls for his valuable comments to
improve this manuscript.

This research has made use of the NASA/IPAC
Extragalactic Database (NED) which is operated by the Jet Propulsion
Laboratory, California Institute of Technology, under contract with the
National Aeronautics and Space Administration.

\bibliography{$HOME/LATEX/refs}
\bibliographystyle{my-mn2e}

\label{lastpage}

\end{document}